\documentclass[aps,prd,preprint,groupedaddress,superscriptaddress, nofootinbib,showkeys]{revtex4-1}
\usepackage{float}
\usepackage{graphicx}
\usepackage{graphics}
\usepackage{epsfig}
\usepackage{latexsym}
\usepackage{colordvi}
\usepackage{amsmath}
\usepackage{amssymb}
\usepackage{slashed}
\usepackage{xcolor}
\usepackage{comment}
\usepackage{romannum}

\usepackage[title]{appendix}
\usepackage{hyperref}

\begin{document}
\pagenumbering{arabic}

\preprint{\parbox[b]{1in}{ \hbox{\tt PNUTP-25/A02}  }}
\preprint{\parbox[b]{1in}{ \hbox{\tt CTPU-PTC-25-28}  }}
\preprint{\parbox[b]{1in}{ \hbox{\tt KIAS-P25032}  }}

\title{Probing the axion-electron coupling at cavity experiments}
\author{Deog Ki Hong}
\email[E-mail: ]{dkhong@pusan.ac.kr}
\affiliation{Department of Physics, Pusan National University,
             Busan 46241, Korea}
            \affiliation{Extreme Physics Institute, Pusan National University, Busan 46241, Korea}
\author{Sang Hui Im}
\email[E-mail: ]{imsanghui@ibs.re.kr}            
\affiliation{Particle Theory and Cosmology Group, Center for Theoretical Physics of the Universe,
Institute for Basic Science (IBS), Daejeon 34126, Korea}
\author{Jinsu Kim}
\email[E-mail: ]{kjs098@ibs.re.kr}
\affiliation{Dark Matter Axion Group, Institute for Basic Science (IBS), Daejeon 34051, Korea} 
\author{TaeHun Kim}
\email[E-mail: ]{gimthcha@kias.re.kr}
\affiliation{School of Physics, Korea Institute for Advanced Study,
Seoul 02455, Korea}    
\author{SungWoo Youn}
\email[E-mail: ]{swyoun@ibs.re.kr}     
\affiliation{Dark Matter Axion Group, Institute for Basic Science (IBS), Daejeon 34051, Korea}

\vspace{0.1in}

\date{\today}

\begin{abstract}
Axion dark matter induces electromagnetic radiation in conductors because of nearly perpetual oscillations of electrons, driven by axion-electron interactions through the so-called chiral magnetic effect. It therefore provides a complementary probe of the axion-electron coupling $g_{ae}$ beyond the conventional axion-photon coupling $g_{a \gamma}$ in cavities. We show that existing axion cavity experiments can constrain the coupling to $g_{ae}\lesssim 10^{-5}$ over the scanned axion mass ranges, $1\,\mu\, {\rm eV}\lesssim m_a\lesssim 20\,\mu\,{\rm eV}$. Although we find that the radiation due to $g_{ae}$ at the copper cavity surface of electric conductivity $\sigma$ is suppressed by $m_a^2/\sigma^2\sim 10^{-20}$, compared to the radiation inside the cavity by the axion-photon conversion due to $g_{a\gamma}$, a sensitivity of about $10^{-9}$  could be achieved for  $g_{ae}$ over a wider range of $m_a$, including values higher than those previously probed, if copper walls are replaced with carbon-based conductors.
\end{abstract}


\maketitle

\newpage

\section{Introduction}
The quantum chromodynamics (QCD)  axion~\cite{Peccei:1977hh,Weinberg:1977ma,Wilczek:1977pj} is a hypothetical but well-motivated particle beyond the Standard Model (SM) of particle physics.  It provides a compelling solution to the strong CP problem and serves as a viable candidate for dark matter (DM), consistent with the standard Big Bang cosmology~\cite{Planck:2018vyg}.

Ever since Sikivie showed over 40 years ago that cosmic DM axions can be detected in a cavity experiment by the axion-photon conversion~\cite{Sikivie:1983ip}, enormous efforts have been put forth to search for axions at  larger scales or with novel detection concepts~\cite{Irastorza:2018dyq}.  Recently, two of the current authors have proposed a new experiment, named as Low temperature Axion Chiral Magnetic Effect (LACME)~\cite{Hong:2022nss}, to measure the axion-electron coupling through chiral magnetic effects (CME) that create a persistent electric current in conductors along external magnetic fields. The measurement of axion-electron coupling could unveil the microscopic origin of axion to  distinguish the axion models such as KSVZ~\cite{Kim:1979if,Shifman:1979if} or DFSZ~\cite{Zhitnitsky:1980tq,Dine:1981rt}. 

In this article, we show that the CME-induced current generates electromagnetic (EM) radiation at the surface of conductors. 
Considering that the cavity walls in existing haloscope experiments such as ADMX and CAPP are made from copper, we derive an upper bound on the axion-electron coupling of $g_{ae}\lesssim 10^{-5}$ for the scanned mass range,  $m_a\sim 1-10~{\rm \mu\, eV}$,  based on current experimental sensitivities~\cite{ADMX:2025vom,CAPP:2024dtx}.   We further discuss the possible enhancement of the CME-induced signal by substituting copper with low-conductivity materials, such as carbon-based conductors, which could potentially improve the sensitivity by several orders of magnitude, reaching $g_{ae} \sim 10^{-9}$ or below~\footnote{Our proposal to use a carbon-based conductor for the cavity wall aims to enhance the sensitivity to the CME signal through a low-cost, minimal  modification of existing cavity experiments, whereas the projection of LACME in~\cite{Hong:2022nss} assumes a maximally ideal setup for measuring the CME signal.}. 
Moreover, since the CME radiation power does not depend on the cavity form factor, higher resonance modes, which generally exhibit higher quality factors, are easily tunable, providing a more efficient probe to higher axion mass regions compared to axion-photon conversion inside the cavity. 

\section{The axion-electron coupling and a momentum shift}
QCD axions couple to gluons through the so-called $\theta$-term to solve the strong $CP$ problem. This makes axions couple to all other SM particles either directly or via loops, enabling various detection strategies for the hypothetical axion DM in our Universe.

If axions constitute most of DM in our universe, 
their occupation number has to be extremely large for axion masses less than a few eV, which 
makes them behave as a coherently oscillating classical field. Since the virialized axion DM in the Milky Way has a velocity dispersion of about $200 \, \text{km}/\text{s}$, we can safely neglect for most of our discussions the gradient term in the classical field representation, whose frequency is localized near the axion mass, $m_a$~\cite{Krauss:1985ub}:
\begin{equation}
a(t)=\sqrt{2T}\int_{-\infty}^{\infty}a(\omega)e^{-i\omega t}\,\frac{d\omega}{2\pi}
\label{dm}
\end{equation}
where $-T\le t\le T$ with  $T\gg m_a^{-1}$\,. The local energy density of axion DM~\cite{Preskill:1982cy,Abbott:1982af,Dine:1982ah} is then 
\begin{equation}
\rho_{\rm DM}\approx\left<\dot a^2\right>=\int_{-\infty}^{\infty}\omega^2\left|a(\omega)\right|^2\,\frac{d\omega}{2\pi}	\approx m_a^2\left<a^2\right>\,,
\end{equation}
where the dot denotes the time derivative. 

As shown in ref.~\cite{Hong:2022nss}, the time derivative of the axion field plays the role of an axial chemical potential $\mu_5$ for electrons through
\!\footnote{The axial number is not conserved for massive electrons. In Fermi liquid like conducting electrons in metal, however, the axial chemical potential controls the total helicity of modes near any point at the Fermi surface, which is conserved up to Adler-Bell-Jackiw anomaly~\cite{Hong:2022nss}.}
\begin{equation}
{\cal L}_{ae}=\frac{g_{ae}}{2m}\,\partial_{\mu}a\,\bar\Psi\gamma^{\mu}\gamma^5\Psi\approx \mu_5\bar\Psi\gamma^0\gamma^5\Psi\,,\quad \mu_5=\frac{g_{ae}}{2m}\,\dot a
\label{ae}
\end{equation}
where $m$ is the electron mass and $g_{ae}$ is the axion-electron coupling. Microscopically, the axion field $a(t)$ or the axial chemical potential $\mu_5$ shifts the electron momentum along its spin direction to result in helicity imbalance.  
To see this, we rewrite the gamma matrices in Eq.~(\ref{ae}) as 
\begin{equation}
\gamma^0\gamma^5=\frac23\vec\gamma\cdot\vec \Sigma\,,
\end{equation}
where $\Sigma^i\equiv \frac{i}{4}\epsilon^{ijk}\gamma^j\gamma^k$ is the spin matrix for the Dirac spinors. The Lagrangian density of free electrons in the background (homogeneous) axion DM becomes then 
\begin{equation}
{\cal L}_{e}=
\bar\Psi\left[\gamma_0\cdot i\partial_t -{\vec\gamma\cdot(i\vec \nabla-\frac23\mu_5\vec\Sigma)}-m\right]\Psi\,
\end{equation}
to show the momentum shift, $\delta \vec p=2\mu_5\vec\Sigma/3$,  to electrons along their spin direction.

The momentum shift results in a macroscopic physical effect on electrons in medium, as it generates a stationary motion such as a persistent electric current in the rest frame of medium~\cite{Hong:2022nss}. 
For instance, if we apply a magnetic field in an electron fluid, all electrons in the lowest Landau level are polarized opposite to the magnetic field direction, getting a net shift along its spin direction by axion DM to create a persistent flow of polarized electrons,  proportional to the momentum shift $\delta\vec p$. This phenomenon is known as chiral magnetic effects that magnetic fields spontaneously create a persistent electric current~\cite{Fukushima:2008xe,Hong:2010hi} 
\begin{equation}
\vec j_{\rm cme}=v_F\frac{e^2}{2\pi^2}\mu_5\vec B_0\,,	
\end{equation}
in a medium of charged fermions having Fermi velocity $v_F$~\cite{Hong:2022nss} with non-vanishing axial chemical potential, $\mu_5$. 
Being a Fermi-surface phenomenon, the CME requires a sufficiently high electron number density for electrons to form a Fermi surface, namely the Fermi energy has to be much larger than the thermal energy, $E_F\gg T$~\cite{Hong:2022nss}.

\section{Axion electrodynamics in a conducting medium}

The axion DM induces an oscillating axial chemical potential $\mu_5$ and hence an oscillating CME current $\vec{j}_{\rm cme}$ in conductors under external magnetic fields. This induces a nearly monochromatic EM radiation at frequency $m_a$, which can be used to detect axion DM.
 
The Lagrangian density for the photon fields, $A_{\mu}$, coupled to axions and an external current, $j_{\mu}$, is given as \begin{equation}
	{\cal L}=-\frac14F_{\mu\nu}F^{\mu\nu}-\frac{{g}_{a\gamma}}{8}a\,\epsilon^{\mu\nu\rho\sigma}F_{\mu\nu}F_{\rho\sigma}\,-j^{\mu}A_{\mu}+{\cal L}_{a}\,,
	\label{lag}
\end{equation}
where the field strength tensor $F_{\mu\nu}=\partial_{\mu}A_{\nu}-\partial_{\nu}A_{\mu}$\, and the axion Lagrangian density  
\begin{equation}
{\cal L}_a=\frac12\partial_{\mu}a\,\partial^{\mu}a-V(a)\,,	
\end{equation}
with the axion potential $V(a)$ due to the explicit breaking of the axion shift symmetry such as the quark mass.
Maxwell's equations become 
\begin{equation}
\partial_{\mu}F^{\mu\nu}+\frac12{g}_{a\gamma}\epsilon^{\mu\nu\rho\sigma}\partial_{\mu}\left(aF_{\rho\sigma}\right)=j^{\nu}\,,	
\end{equation}
supplemented with the usual Bianchi identity
\begin{equation}
\epsilon^{\mu\nu\rho\sigma}\partial_{\nu}F_{\rho\sigma}=0\,.	
\end{equation}
In terms of the electric and magnetic fields, $E^i=F^{i0}$ and $B^i=\frac12\epsilon^{ijk}F_{jk}$\,, we have 
\begin{eqnarray}
\vec\nabla\cdot\left(\vec E-{\bar g}_{a\gamma}a\vec B\right)=j^0\,,\quad 
\vec\nabla\times\left(\vec B+{g}_{a\gamma}a\vec E\right)-\frac{\partial}{\partial t}\left(\vec E-{g}_{a\gamma}a\vec B\right)=\vec j\;	\label{modified}\\
\vec\nabla\cdot\vec B=0\,,\quad \vec\nabla\times\vec E+\frac{\partial}{\partial t}\vec B=0\,,
\label{bianchi}
\end{eqnarray}
where we have taken both the dielectric constant and magnetic susceptibility of conductors to be those of vacuum. 

Now, we assume that the axion field or its time derivative are homogeneous under an external magnetic field, $\vec B_0(\vec x)$, with no net charge density, $j^0=0$.
As the axion couples to the photon fields, the background axion fields will source the EM fields.
Since ${g}_{a\gamma}$ or $g_{ae}$ are quite small compared to the natural scale of the problem, we solve Maxwell's equations up to the linear order in ${g}_{a\gamma}$ or $g_{ae}$ to find the induced EM fields by axions:
\begin{equation}
\vec B=\vec B_1+\vec B_0,\quad \vec E=\vec E_1, \quad 	\vec j=\vec j_1+\vec j_{\rm cme}\,,
\end{equation}
where all the quantities with the subscript $1$ are of order ${g}_{a\gamma}$ or $g_{ae}$ and the CME current, $\vec j_{\rm cme}$, exists only inside the conductor. 
In conductors, the induced current $\vec{j}_1$ is approximately ohmic when the frequency of the field is small compared to the conductivity $\sigma$
\,\footnote{For copper at room temperature, $\sigma=6\times 10^7\,{\rm S/m}$, and $1\,{\rm S/m}\simeq1.13\, \times 10^{11}\,{\rm s}^{-1}$ in natural units.}, namely
\begin{equation}
\vec j_1=\sigma \vec E_1\,
\end{equation}
for $\omega\ll\sigma$. 
Maxwell-Ampere's law in Eq.~(\ref{modified}) then becomes 
\begin{equation}
\vec\nabla\times \vec B_1-\frac{\partial \vec E_1}{\partial t}=\sigma\vec E_1-{\bar g}_{a\gamma}\dot a\vec B_0\,,
\label{amp}
\end{equation}
where ${\bar g}_{a\gamma}\equiv g_{a\gamma}+(\alpha/{\pi})\, v_Fg_{ae}/m$ to include the CME current on top of the vacuum current, $\vec j_{\rm vac}=g_{a\gamma}\dot a\vec B_0$, coming from the anomalous axion-photon coupling in Eq.~(\ref{lag}).
Taking the time derivative of Maxwell-Ampere's law, we obtain 
\begin{equation}
\vec\nabla\times\frac{\partial \vec B_1}{\partial t}-\frac{\partial^2\vec E_1}{\partial t^2}=\sigma\frac{\partial \vec E_1}{\partial t}-{\bar g}_{a\gamma}{\ddot a}\vec B_0\,.	
\end{equation}
Utilizing Faraday's law 
together with Gauss's law, we find the wave equation in a conducting medium, sourced by the magnetic field, 
\begin{equation}
\nabla^2\vec E_1-\frac{\partial^2\vec E_1}{\partial t^2}=\sigma\frac{\partial \vec E_1}{\partial t}-{\bar g}_{a\gamma}{\ddot a}\vec B_0\,.	
\label{wave}
\end{equation}
The source for the EM waves exists whenever $\ddot a\vec B_0$ is non-vanishing. Since, however,  we are interested in the CME current in the conductor, we constrain the source to be the one inside the conductor.\,\footnote{The source outside conductor will create electric fields inside the conductor but highly screened beyond the penetration depth.} Namely, we take   $\vec B_0(\vec x)$ in Eq.~(\ref{wave}) to be $\vec B_0(\vec x)\,\theta(\vec x)$ with a step function $\theta(\vec x)=1$ inside the conductor and zero outside.
With the Fourier transform, $\vec E_1(\vec x,t)=\sqrt{2T}\int_{-\infty}^{\infty}\vec E_m(\vec x,\omega)e^{-i\omega t}{d\omega}/2\pi$, one obtains the amplitude in the frequency space as 
\begin{equation}
\vec E_m(\vec x,\omega)=\vec E_{m,0}(\vec x,\omega)+{\bar g}_{a\gamma}\omega^2a(\omega)
\int_q
\frac{\vec B_0(q)\,e^{i\vec q\cdot \vec x}}{\omega^2+i\omega\sigma-q^2}
\end{equation}
where $\vec B_0(q)$ is the spatial Fourier transform of the magnetic field inside the conductor, $\vec B_0(\vec x)\,\theta(\vec x)$, and $\vec E_{m,0}$ is the homogeneous solution that  satisfies
$\left(\nabla^2+\omega^2+i\omega\sigma\right)\vec E_{m,0}=0$\,.

\subsection{Infinite conductor}
\begin{figure}[th]
\centering 
\includegraphics[scale=0.32]{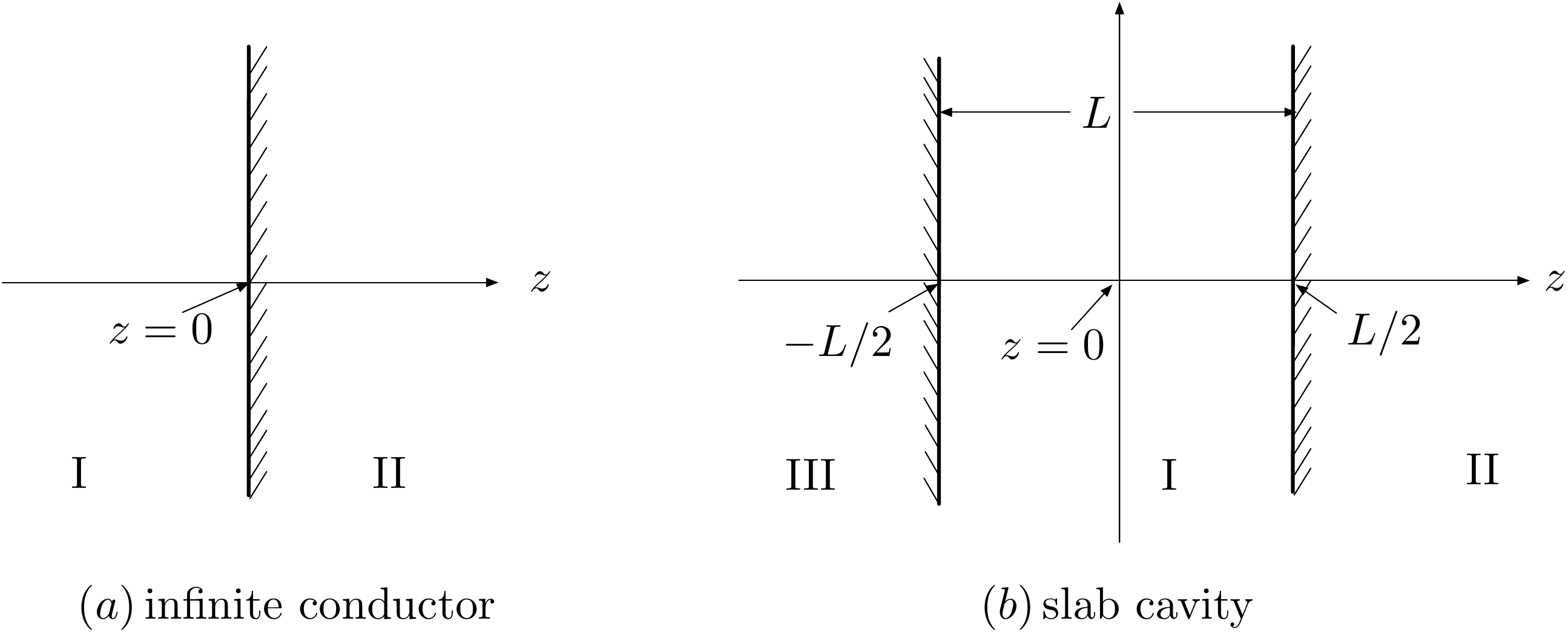}
\caption{(a) An infinite conductor occupying Region $\rm II$ ($z\ge0$). (b) A cavity formed by two parallel infinite conductors, separated by a distance $L$.}
\label{cond}
\end{figure}

Let us now consider an infinite conductor occupying the region  $z\ge0$, Region $\rm II$ in Fig.~\ref{cond} (a), with its surface being the $xy$-plane. 
We further assume, for simplicity, that the background magnetic field is applied only inside the conductor and is perpendicular to the $z$ direction, $\hat z\cdot\vec B_0=0$\,.
Then, for a plane EM wave, propagating along the $z$ direction,  the amplitude of the electric field can be written as
\begin{equation}
\vec E_{m,0}(\omega,z)=\vec D(\omega)\,e^{ikz}+\vec F(\omega)\,e^{-ikz}\,.	
\end{equation}
Inside the conductor, the wave vector satisfies 
\begin{equation}
-k^2+\omega^2+i\omega\sigma=0\quad {\rm or}\quad k=\sqrt{\omega}\left(\omega^2+\sigma^2\right)^{1/4}e^{i\phi/2}\,,	
\end{equation}
where $\phi=\tan^{-1}\left(\sigma/\omega\right)$. 
Since we do not have EM waves propagating from $z=\infty$, 
$\vec F=0$.
The amplitude of electric fields in the conductor is then given as
\begin{equation}
\vec E_m(\omega,z)=\vec D(\omega)\,e^{-z/\delta}e^{i{\rm Re}[k]z}
+{\bar g}_{a\gamma}\omega^2a(\omega)\int_{-\infty}^{\infty}\frac{\vec B_0(q)\,e^{iqz}}{\omega^2+i\omega\sigma-q^2}\frac{dq}{2\pi}\,,
\label{electric1}
\end{equation}
where $\delta$ is the penetration depth for EM waves of frequency $\omega$, given as~\footnote{The screening of axion-induced currents in medium under an external magnetic field has been widely studied~\cite{Slonczewski:1985oco,Mitridate:2020kly,Marsh:2022fmo,Berlin:2023ppd}. Our result, Eq.~(\ref{electric1}), agrees with, for example, ref.~\cite{Marsh:2022fmo}, which is, though,  more focused  on the resonance with quasi-particles in dielectric materials, not in cavity. See also ref.~\cite{Berlin:2023ubt}, which considers radiation induced by the axion-electron coupling from bound electrons in dielectric materials.} 
 
\begin{equation}
\delta=\left({\rm Im}[k]\right)^{-1}=\sqrt{\frac{2}{\omega\sigma}}
\left[\frac{\omega}{\sigma}+\sqrt{\frac{\omega^2}{\sigma^2}+1}\,\right]^{1/2}\,.\end{equation}

Plugging Eq.~(\ref{electric1}) into Maxwell-Ampere's law, Eq.~(\ref{amp}), with the frequency amplitude $\vec B_m(\omega,z)$ of $\vec B_1$, we find 
\begin{equation}
\vec\nabla\times\vec B_m=\left(\sigma-i\omega\right) \vec E_m(\omega,z)+i{\bar g}_{a\gamma}\omega a(\omega)\vec B_0\,.\label{amp2}	
\end{equation}
We see that the propagating component of magnetic field, $\vec B_1$, generated by the axion CME or vacuum current decays  in conductors over the penetration depth, $\delta$, as well. 
One can also derive the induced magnetic field from Faraday's law, which gives at the leading order in ${\bar g}_{a\gamma}$
\begin{equation}
\vec B_m(\omega,z)=\frac{k}{\omega}\hat z\times \vec D\,e^{ikz}+{\bar g}_{a\gamma}\omega a(\omega)\int_{-\infty}^{\infty}\frac{\hat z\times\vec B_0(q)q\,e^{iqz}}{\omega^2+i\omega\sigma-q^2}\frac{dq}{2\pi}\,.	
\label{magnetic}
\end{equation}
Using $\hat z\cdot \vec D=0=\hat z\cdot \vec B_0$ and $k^2=\omega(\omega+i\sigma)$,  one can check that this solution recovers Eq.~(\ref{amp2}) to be consistent with  Maxwell's equations.

In Region $\rm I$ (the free space) of Fig.~\ref{cond}\,(a),  where there is no source or $\vec B_0=0$, there will be, however,  traveling EM waves in general. They can be written as, assuming no incoming waves from the far left, 
\begin{equation}
\vec E(z,t)=\sqrt{2T}\int_{-\infty}^{\infty}\vec C(\omega)\,e^{-i(pz+\omega t)}\frac{d\omega}{2\pi}\,,\,
\vec B(z,t)=-\sqrt{2T}\int_{-\infty}^{\infty}\frac{p}{\omega}\hat z\times \vec C\,e^{-i(pz+\omega t)}\frac{d\omega}{2\pi}\,.
\end{equation}
When there are no surface charges or currents, both 
electric fields and magnetic fields have to be continuous at the surface of the conductor~\cite{landau8}, assuming the magnetic permeability of the conductor $\mu_m=1$.  The continuity of the electric field gives
 \begin{equation}
\vec C=\vec D+{\bar g}_{a\gamma}\omega^2a(\omega)\int_{-\infty}^{\infty}\frac{\vec B_0(q)}{\omega^2+i\omega\sigma-q^2}\frac{dq}{2\pi}\,,
\end{equation}
and that of the magnetic fields gives with the vacuum dispersion relation $\omega= p$,  
\begin{equation}
-\hat z\times \vec C\,=\frac{k}{\omega}\hat z\times \vec D+{\bar g}_{a\gamma}\omega a(\omega)\int_{-\infty}^{\infty}\frac{\hat z\times\vec B_0(q)q}{\omega^2+i\omega\sigma-q^2}\frac{dq}{2\pi}\,.
\end{equation}
Since both the propagating EM fields and $\vec B_0$ are perpendicular to the propagating direction, we get from the two continuity conditions 
\begin{equation}
\vec C(\omega)={\bar g}_{a\gamma}\frac{\omega^2a(\omega)}{\omega+k}\int_{-\infty}^{\infty}\frac{\vec B_0(q)}{q+k}\cdot \frac{dq}{2\pi}\,,\quad 
\vec D(\omega)={\bar g}_{a\gamma}\frac{\omega^2 a(\omega)}{\omega+k}\int_{-\infty}^{\infty}\frac{\vec B_0(q)(q+\omega)}{q^2-k^2}\cdot\frac{dq}{2\pi}\,.                                     
\end{equation}
When the external magnetic field $\vec B_0(z)=\vec B_0$ is constant, the Fourier transform of the magnetic field in the conductor $\vec B_0\,\theta (z)$ becomes with $\epsilon\to0^+$ 
\begin{equation}
\vec B_0(q)=	-i\frac{\vec B_0}{q-i\epsilon}\,.
\end{equation}
The amplitudes of the EM fields become
\begin{equation}
\vec C(\omega)=i{\bar g}_{a\gamma}a(\omega)\vec B_0(-k)\frac{\omega^2}{\omega+k},\quad \vec D(\omega)=-i{\bar g}_{a\gamma}a(\omega)\vec B_0(-k)\frac{\omega^2(\omega-k)}{2k(\omega+k)}	\,.
\end{equation}
For $\sigma\gg \omega$, one finds 
\begin{equation}
\vec C(\omega)\simeq i{\bar g}_{a\gamma}\vec B_0\frac{\omega a(\omega)}{\sigma}\simeq 2\vec D(\omega)\,.
\end{equation}

We see that a conductor under constant magnetic field will emit EM waves at its surface if there is axion DM with a non-vanishing time derivative. 
The mean value of the Poynting vector emitted by the conductor is given as for $\sigma\gg m_a$ 
\begin{equation}
\vec S=\frac{1}{2}{\rm Re}\left[\vec E\times \vec B^*\right]\simeq -\frac{{\bar g}^2_{a\gamma}\left<{\dot a}^2\right>}{2\sigma^2}B_0^2\,\hat z\,.	
\label{power}
\end{equation}
The power emitted by the conductor of surface area $A$ is then estimated to be 
\begin{equation}
P_a=9.5\times 10^{-34}\,{\rm W}\left(\frac{10^{14}\,{\rm GeV}}{{\bar g}^{-1}_{a\gamma}}\right)^2\left(\frac{\left<\dot a^2\right>}{0.45\,{\rm GeV/cm^3}}\right)\left(\frac{10^{14}\,\rm s^{-1}}{\sigma}\right)^2\left(\frac{B_0}{10\,{\rm T}}\right)^2\left(\frac{A}{1\rm m^2}\right)\,,	
\end{equation}
where, for our estimate, we take a carbon-based conductor whose conductivity is a few thousand times smaller than that of copper~\cite{pierson}.

\subsection{Resonant Cavity}

The axion cavity experiments such as ADMX or CAPP rely on the axion-photon conversion under an external magnetic field inside the cavity. When the frequency of the cavity normal modes is tuned at the axion mass, the conversion signal is resonantly enhanced for the axion detection. We note in the cavity experiments that, as the applied magnetic fields are also present in the cavity wall, there should be CME currents at the wall, flowing along the applied magnetic field. The cavity wall therefore radiates CME-induced  EM waves.

Inside the cavity, there will be emitted waves from all the points of the wall to interfere, forming standing waves at the resonance. 
To study the cavity resonance of EM waves emitted from the wall, we introduce, for simplicity, two parallel infinite conductors separated by a
distance $L$ so that the cavity becomes an infinite slab, Region \Romannum{1} of Fig.~\ref{cond}\,(b).

In Region \Romannum{2} of Fig.~\ref{cond}\,(b), the EM fields will be the same as in the case of an infinite conductor, Eqs.~(\ref{electric1}) and (\ref{magnetic}), except that the amplitudes are now modified due to the presence of incidents waves:
\begin{equation}
\vec E_2(\omega,z)=\vec D_2(\omega)\,e^{ikz}
+{\bar g}_{a\gamma}\omega^2a(\omega)\int_{-\infty}^{\infty}\frac{\vec B_0(q)\,e^{iqz}}{\omega^2+i\omega\sigma-q^2}\frac{dq}{2\pi}\,,	
\end{equation}
where the left-moving waves are absent as before since there are no sources from the far right.
Henceforth, we take ${\bar g}_{a\gamma}=v_F(\alpha/\pi)g_{ae}/m$, putting $g_{a\gamma}=0$, to isolate the radiation from the CME currents. 
On the other hand, in Region \Romannum{1} of Fig.~\ref{cond}\,(b)\,\footnote{The incoming and outgoing waves are coherent and therefore interfere, since the coherence length of axion DM, given by its de Broglie wave length, is much larger than the size of cavity. }, we have both incoming and outgoing waves, respecting the symmetry of cavity under $z\to-z$, 
\begin{equation}
\vec E(\omega,z)=\vec C_1(\omega)\,e^{-i\omega z}\pm\vec C_1(\omega)\,e^{i\omega z}\,.
\label{cavity_e}
\end{equation}
Requiring the continuity of both electric and magnetic fields, we find for the positive parity fields, which are even under the reflection, $z\to-z$, 
\begin{eqnarray}
\vec C_1(\omega)	&=&\frac{{\bar g}_{a\gamma}\omega a(\omega)}{4\cos(\omega L/2)} \vec B_0(-k)e^{-ikL/2}\left[F(\omega)-i\frac{\omega}{k}\right]\,\\
\vec D_2(\omega)&=&\frac12{\bar g}_{a\gamma}\omega a(\omega) \vec B_0(-k)e^{-ikL}F(\omega)\,,
\end{eqnarray}
where 
\begin{equation}
F(\omega)=\frac{\cos(\omega L/2)+i\frac{\omega}{k}\sin(\omega L/2)}{\sin(\omega L/2)+i\frac{k}{\omega}\cos(\omega L/2)}\,.	
\end{equation}
If a constant magnetic field, $\vec B_0$, is  applied only to the cavity wall, $|z|\ge L/2$, to measure the radiation by the CME currents, 
the positive-parity electric field in the cavity becomes
\begin{equation}
\vec E(\omega,z)	\simeq\frac{{\bar g}_{a\gamma}\omega a(\omega)}{\cos(\omega L/2)\sigma}\,\vec B_0\cos(\omega z)
\end{equation}
and the negative-parity electric field, odd under the reflection, becomes
\begin{equation}
\vec E(\omega,z)\simeq	\frac{{\bar g}_{a\gamma}\omega a(\omega)}{\sin(\omega L/2)\sigma}\,\vec B_0\sin(\omega z)\,
\end{equation}
in the large conductivity limit $\sigma\gg\omega$.

\begin{figure}[h]
\centering 
\includegraphics[scale=0.38]{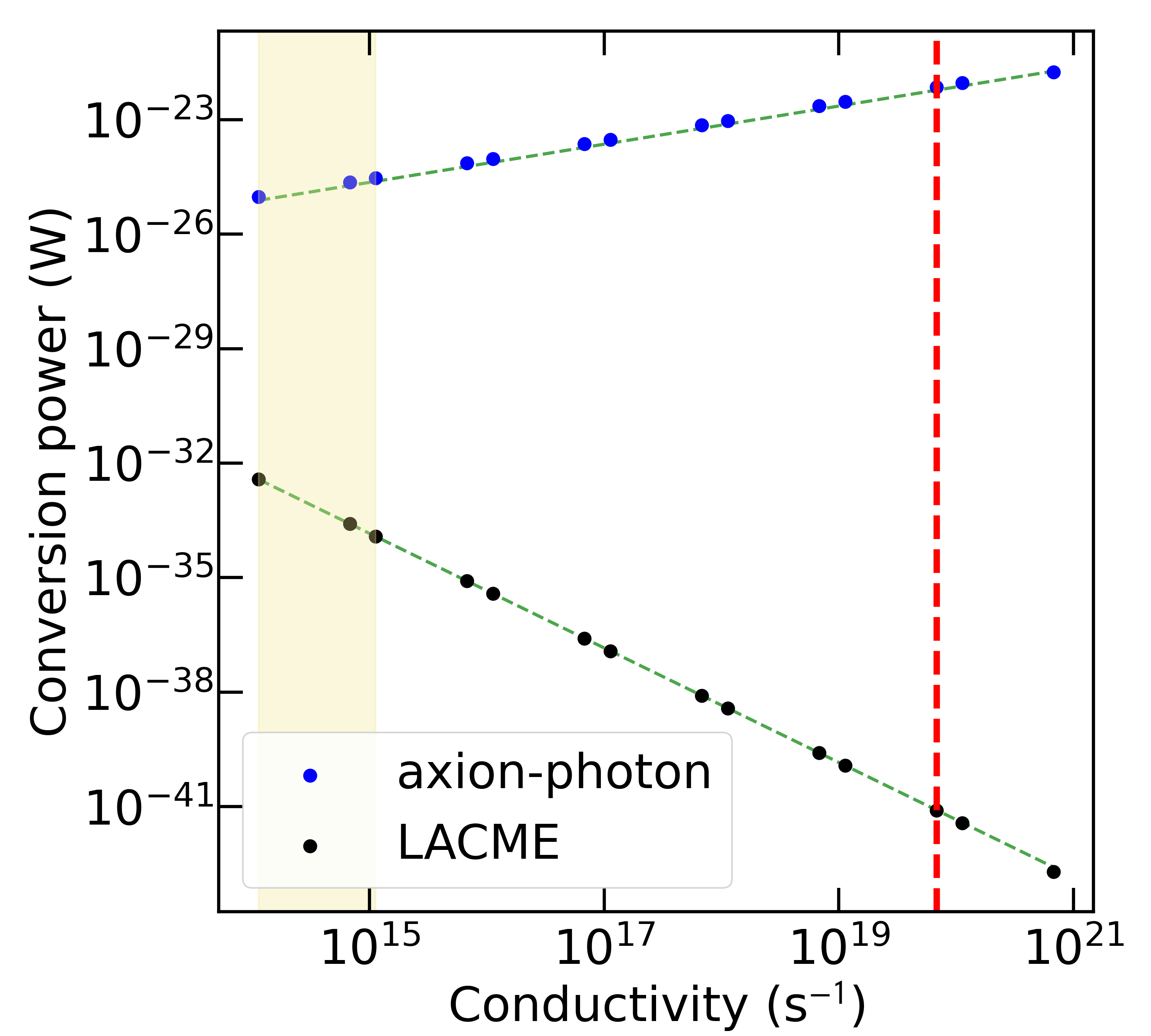}
\caption{Numerically computed radiation power of the ${\rm TM}_{010}$ mode for a cylindrical cavity with diameter 90\,mm and height 180\,mm in a $10\,{\rm T}$ magnetic field with $g_{a\gamma}$ or ${\bar g}_{a\gamma}$ at $10^{-14}\,{\rm GeV}^{-1}$. 
The blue dots represent the conventional axion-photon conversion and the black ones denote the CME-induced radiation.
The green dashed lines indicate fitted curves following the relations \(P_a \propto \sigma^{0.5}\) and \(P_a \propto \sigma^{-1.5}\) for the two cases. 
The red vertical line marks the conductivity of copper at cryogenic temperatures, while the yellow band represents the conductivity range of conductive carbon. }
\label{comsol}
\end{figure}

If we expand the electric fields, Eq.~(\ref{cavity_e}), in terms of cavity modes~\cite{Krauss:1985ub},
\begin{equation}
\vec E(\omega,z)=\sum_{j=0}^{\infty}\left[\lambda^+_j(\omega)\vec e_{j+}(z)+\lambda^-_j(\omega)\vec e_{j-}(z)\right]\,,	
\end{equation}
where $\vec e_{j+}(z)=\hat e \cos(\omega_jz)$ is the positive-parity cavity mode with eigen frequencies $\omega^+_j=(2j+1)\pi/L$ and the polarization vector $\hat e$, while $\vec e_{j-}(z)=\hat e \sin(\omega_jz)$ being the negative-parity mode with $\omega^-_j=2j\pi/L$. 
If the radiated waves in the cavity resonate with $j$-th positive-parity mode, $\omega\approx\omega^+_j$, 
the amplitude of the cavity mode becomes near the resonance
\begin{equation}
\lambda^+_j(\omega)\simeq \frac{2{\bar g}_{a\gamma}\omega a(\omega)}{\cos(\omega L/2)\sigma L}{\hat e}\cdot \vec B_0\cdot \frac{\sin\left[(\omega-\omega_j)L/2\right]}{\omega-\omega_j}\,.
\end{equation}
One gets a similar expression for the negative-parity electric fields to find $|\lambda_j^+(\omega)|=|\lambda_j^-(\omega)|$ near the resonance.
Introducing the loaded quality factor, $Q$, of the cavity, the mode energy stored in the cavity of surface area $A$ becomes 
\begin{equation}
U_j\simeq\frac{2{\bar g}_{a\gamma}^2A}{ \sigma^2L}(\hat e\cdot\vec B_0)^2\int_{-\infty}^{\infty}\frac{\left|a(\omega)\right|^2\omega^2}{(\omega-\omega_j)^2+\omega^2/Q^2}\frac{d\omega}{2\pi}\approx \frac{2{\bar g}_{a\gamma}^2|\vec B_0|^2}{\sigma^2}\frac{\rho_{\rm DM}V
}{(m_aL)^2}Q^2\,,
\end{equation}
where we used the cavity volume $V=AL$ and the polarization $\hat e\parallel \vec B_0$. 
Compared with the axion-photon conversion inside the cavity, the mode energy stored in the cavity is suppressed by $m_a^2/\sigma^2$. 
Since the power absorbed by the cavity near the resonance ($\omega_j\approx m_a$) is $P_a\approx U_j\omega/Q$,  
the power due to the CME currents from the cavity wall becomes 
\begin{equation}
P_a\approx \frac{{\bar g}_{a\gamma}^2|\vec B_0|^2}{\sigma^2}\frac{\rho_{\rm DM}A}{\gamma_j}\,Q,	
\end{equation}
which agrees with the radiation power at the surface, Eq.~(\ref{power}), taking into account the interference, up to the $Q$ factor and the cavity geometry factor, $\gamma_j=m_a L$ for the slab cavity. For a long cylindrical cavity of radius $R$, the geometric factor $\gamma_0=0.35\,(m_aR)^3$ and $m_a=2.41\,R^{-1}$ at the lowest normal mode or ${\rm TM}_{010}$ mode to give the power to be  
\begin{equation}
\!\!\!P_a=1.7\times 10^{-41}\,{\rm W}	\!\left(\frac{10^{14}\,{\rm GeV}}{{\bar g}^{-1}_{a\gamma}}\right)^2\!\left(\frac{\left<\dot a^2\right>}{0.45{\rm GeV}/{\rm cm}^{3}}\right)\!\left(\frac{10^{20}{\rm s}^{-1}}{\sigma}\right)^2\!\left(\frac{B_0}{10\,{\rm T}}\right)^2\left(\frac{A}{1\rm m^2}\right)\left(\frac{Q}{4.5\times10^4}\right)\,.	
\end{equation}
Since the quality factor is proportional to $\sqrt{\sigma}$, the absorbed power by the CME currents at the wall scale as $P_a\sim \sigma^{-1.5}$, which has been checked for a cylindrical cavity by the numerical simulation,  using COMSOL~\cite{comsol6.2}, shown in Fig.~\ref{comsol}~\footnote{In experiments, one needs additionally to insert a dielectric or conducting rod   inside the cavity to scan the axion mass. We find by numerical simulations that the effect of a rod does not alter the order of radiation power.  }.   

\begin{figure}[H]
\centering 
\includegraphics[scale=0.6]{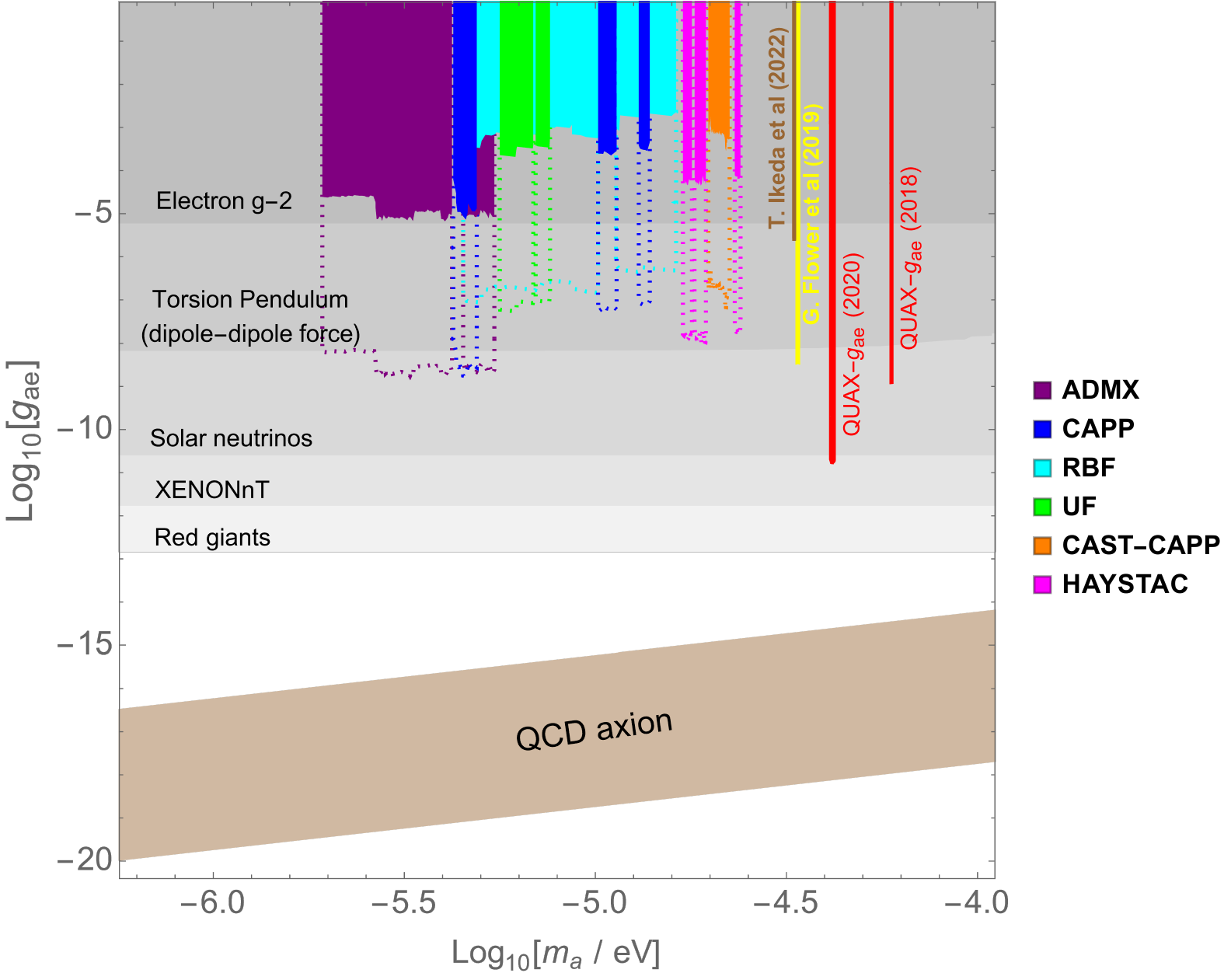}
\caption{Current and projected constraints on the axion-electron coupling $g_{ae}$. The color-shaded regions are excluded by existing data from the corresponding cavity experiments: ADMX~\cite{ADMX:2025vom,ADMX:2024xbv,ADMX:2021nhd,ADMX:2019uok,ADMX:2018gho,ADMX:2009iij}, CAPP~\cite{Bae:2024kmy,CAPP:2024dtx,Kim:2023vpo,Yang:2023yry,Yi:2022fmn,Kim:2022hmg,Lee:2022mnc,CAPP:2020utb,Jeong:2020cwz,Lee:2020cfj}, RBF~\cite{DePanfilis:1987dk}, UF~\cite{Hagmann:1990tj}, CAST-CAPP~\cite{Adair:2022rtw}, and HAYSTAC~\cite{HAYSTAC:2024jch,HAYSTAC:2020kwv,Brubaker:2016ktl}. The dotted lines represent the projected sensitivity of these experiments assuming carbon-based cavities, obtained from the power ratio (Fig.~\ref{comsol}) between the axion-photon conversion and the CME-induced radiation, which does not change much by the details of each cavity experiments. The narrow red band shows the exclusion limit set by QUAX~\cite{Crescini:2018qrz,QUAX:2020adt}, while the narrow yellow and brown bands are from the exclusion by~\cite{Flower:2018qgb} and~\cite{Ikeda:2021mlv}, respectively. The gray shaded regions indicate exclusions from other experiments and astrophysical constraints~\cite{Yan:2019dar,Terrano:2015sna,Gondolo:2008dd,XENON:2022ltv,Capozzi:2020cbu} and the brown band represents the QCD axion prediction.}
\label{sensitivity}
\end{figure}

Finally, existing bounds on $g_{a \gamma}$ can be translated into bounds on $g_{ae}$. For instance, the current sensitivity of CAPP, ${g}_{a\gamma}<10^{-15}\,{\rm GeV^{-1}}$ at $m_a = 1 \, \text{GHz}$~\cite{CAPP:2024dtx}, gives $g_{ae}<10^{-5}$ with $v_F=10^{-2}$.  The constraints on $g_{ae}$ from other experimental or observational data are shown in Fig.~\ref{sensitivity}. 

\subsection{Cavity design for LACME}

Once the radiation emitted at the cavity wall by the CME currents resonates within the cavity, the radiation energy will accumulate inside the cavity to enhance the CME signal. There are, however, two limitations to this enhancement. The first one is that the axions are virialized, having a width in the velocity distribution or a dispersion in energy distribution, $\Delta E$. The quality factor for virialized axions has been estimated to be~\cite{Krauss:1985ub} 
\begin{equation}
	Q_a\equiv\frac{m_ac^2}{\Delta E}\sim 10^6\,.
\end{equation}
The other limitation comes from the fact that the cavity wall is not a perfect reflector, losing some of radiation energy upon reflection. Such intrinsic quality factor of the cavity 
is given as the ratio of the incident energy at the wall over the transmitted energy, which is determined by the conductivity of the cavity wall and the radiation frequency as 
\begin{equation}
Q_c\equiv\frac{{\cal E}_I}{\Delta {\cal E}}\sim \sqrt{\frac{\sigma}{m_a}}\,.	
\end{equation}
The intrinsic quality factor is typically smaller than the axion quality factor and thus defines the quality factor of the axion cavity experiments in general. 

Since the radiation power of the CME current is proportional to $\sigma^{-1.5}$ as long as $m_a\ll \sigma$, it is beneficial to use a poor conductor for the cavity wall\,\footnote{Any conductors will exhibit chiral magnetic effects as long as they have Fermi surfaces.}, even at the cost of a reduced quality factor.   For example, we expect to achieve $g_{ae}<10^{-9}$ by replacing the copper wall of the cavity with a carbonic conductor in the current setup at CAPP (see the plot shown in Fig.~\ref{sensitivity}).~\footnote{Recently, an interesting experiment (MOSAIC) has been proposed to search for the spin-dependent interactions of electron-coupled axion dark matter~\cite{Chang:2025sno}. }
Furthermore, as the radiation is sourced by the CME current at the wall, the power is not reduced by the cavity form factor, which enables us to better tune higher resonance modes that generally exhibit higher $Q$-factors. One can therefore easily probe the axion-electron coupling at higher axion mass in cavity experiments.

\section{Conclusion}

To conclude, we have solved Maxwell's equations for a conducting medium with a boundary under the homogeneous axion DM and an external magnetic field, showing that the conductors radiate EM  waves sourced by CME currents.  
The radiated waves can be resonantly stored in cavities, leading a signal enhancement. In consequence, we find that the existing cavity experiments such as ADMX or CAPP places constraints on the axion-electron coupling at the level of $g_{ae}<10^{-5}$\,.
Compared with the radiation from the axion-photon conversion inside the cavity, the CME radiation is suppressed by $m_a^2/\sigma^2$ for cavities of good conductors, $\sigma\gg m_a$. We estimate, however, that the axion-electron coupling up to $10^{-9}$ can be probed by replacing the cavity wall with low-conductivity material such as carbon, with similar or better reach at higher axion mass regions. Finally, we emphase that, if an axion signal is observed by cavity experiments, one has to repeat the experiment again to pin down the correct source for the signal in the apparatus by turning off the external magnetic field inside the cavity, while keeping the magnetic field in the wall. If there is no significant change in the signal, it is very likely due to the CME current.


\acknowledgments

We are grateful to Pierre Sikivie for the useful comments. 
DKH also expresses his gratitude to Hyesung Kang for her hospitality during the visit, and to Jeff Dror, Kent Irwin, and Alex Shushkov for valuable discussions at the Axions 2024 conference, Gainesville, Florida, U.S.A. 
This research was supported by Basic Science Research Program through the National Research Foundation of Korea (NRF) funded by the Ministry of Education (NRF-2017R1D1A1B06033701) (DKH), and IBS under the project codes, IBS-R018-D1 (SHI) and IBS-R040-C1 (JSK and SWY). THK is supported by KIAS Individual Grant PG095201 at Korea Institute for Advanced Study.



\begin{thebibliography}{99}
\bibitem{comsol6.2}
The simulations were performed using COMSOL Multiphysics® software, version 6.2.
\end{thebibliography}
\end{document}